# Fitting the Log Skew Normal to the Sum of Independent Lognormals Distribution

## Marwane Ben Hcine[1] and Ridha Bouallegue[2]

[1,2]Innovation of Communicant and Cooperative Mobiles Laboratory, INNOV'COM
Sup'Com, Higher School of Communication
Univesity of Carthage
Ariana, Tunisia
[1]marwen.benhcine@supcom.tn
[2]ridha.bouallegue@supcom.rnu.tn

## ABSTRACT

*Sums of lognormal random variables (RVs) occur in many important problems in wireless communications especially in interferences calculation. Several methods have been proposed to approximate the lognormal sum distribution. Most of them requires lengthy Monte Carlo simulations, or advanced slowly converging numerical integrations for curve fitting and parameters estimation. Recently, it has been shown that the log skew normal distribution can offer a tight approximation to the lognormal sum distributed RVs. We propose a simple and accurate method for fitting the log skew normal distribution to lognormal sum distribution. We use moments and tails slope matching technique to find optimal log skew normal distribution parameters. We compare our method with those in literature in terms of complexity and accuracy. We conclude that our method has same accuracy than other methods but more simple. To further validate our approach, we provide an example for outage probability calculation in lognormal shadowing environment based on log skew normal approximation.*

## 1. INTRODUCTION

The sum of Lognormals (SLN) arises in many wireless communications problems, in particular in the analyses the total power received from several interfering sources. With recent advances in telecommunication systems (Wimax, LTE), it is highly desired to compute the interferences from others sources to estimate signal quality (*Qos*). Such interference can be modelled as sum of lognormal distribution. Several approaches were proposed to obtain the resulting sum of lognormal distributions. Approximation methods can be categorized into three main classes: lognormal approximations [2-6], numerical methods [7-10] and closed form approximations methods [11-15]. Lognormal approximations methods are based on the fact that lognormal sum can be modelled by another lognormal distribution; these methods are simple yet not accurate particularly in upper and lower regions. Numerical methods are accurate but it need extensive numerical integrations. Closed form approximations methods are generally not accurate at some regions .i.e. does not catch the whole body of the SLN distribution.

Furthermore, the main drawback of almost all of these methods is the need for an extensive Monte Carlo simulation in order to find optimal parameters of the proposed distribution. Except from [2], [12] and [16], all closed form approximations methods need a prior knowledge of the SLN *cdf* for curve fitting and parameters estimation. In [2], a simple lognormal approximation of the SLN is proposed. Using Moment Matching Method (MOM), distribution parameters could be estimated yet this approximation is not accurate in upper and lower regions of the *cdf*.

In [12], a Type IV Pearson Approximation method is proposed. For parameters estimation, they used moment matching in the logarithm domain which is handy to implement. This method has good accuracy in a wide region, but it did not offer good approximations all the cases (i.e. large spread power $\sigma_j$). In [16], a Modified–Power–Lognormal (MPLN) approximation method is used. Distribution parameters are estimated using some properties of the upper and lower tails of the SLN distribution [17]. It is easy to implement but it did not offer a tight approximation of the SLN distribution in all the cases (different/lower spread power $\sigma_i$).

Recently, it has been shown that Log Skew Normal distribution (LSKN) could approximate the SLN distribution over a wide range. In [18] Wu and Li showed that Log Skew Normal provide a very tight approximation to the SLN distribution. In [19], they proposed a transformed LSN to control the skewness of samples in the transform logarithm domain. However, to estimate distribution parameters, they used in both approximations the moment matching method in the logarithm domain, which require prior knowledge of the SLN distribution.

In this paper, we propose a simple method based on moment matching method (in linear domain) and tails slope match of both SLN and LSKN distributions. Given a set of pairs $\{(\mu_i, \sigma_i^2)\}_{i=1}^N$, the goal of our work is to estimate optimal parameters for log skew normal distribution. We derive the expression of the first two central moments of SLN and LSKN distributions. Using moments and tails slope matching, we provide an accurate approximation to optimal parameters values for LSKN distribution. We validate our approach by an example for outage probability calculation in lognormal shadowing environment.

The rest of the paper is organized as follows: In section 2, a brief description of the lognormal and sum of lognormal distribution is given. Moment and tails properties of SLN distribution are investigated. In section 3, we introduce the log skew normal distribution and its parameters. Moment and tails properties of the LSKN distribution are studied, then, we use MoM and tails slope matching to estimate distribution parameters. In section 4, we compare our method with moment matching method in logarithm domain [18-19]. Also, we provides comparisons with some existing approximation method (Lognormal approximation and Modified Power Lognormal) with the simulation results on the cumulative distribution function *(CDF)* of sum of *N* lognormal random variables in different conditions. In section 5, we give an example for outage probability calculation in lognormal shadowing environment based on our method. The conclusion remarks are given in Section 6.

## 2. LOGNORMAL SUM DISTRIBUTION

### 2.1. Lognormal PDF

Let $X$ be a RV with a normal distribution then $L = e^X$ has a lognormal distribution. Likewise, if $L$ is lognormally distributed, then ln(L) is normally distributed. Given $X$, a Gaussian RV with mean $\mu_X$ and variance $\sigma_X^2$, $L = e^X$ is a lognormal RV with (PDF):

$$f_L(l, \mu_X, \sigma_X) = \begin{cases} \dfrac{1}{\sqrt{2\pi}l\sigma_X}\exp(-\dfrac{1}{2\sigma_X^2}\big[\ln(l) - \mu_X\big]^2) & l > 0 \\ 0 \quad otherwise \end{cases} \quad (1)$$

$$= \phi(\frac{\ln(l) - \mu_X}{\sigma_X})$$

Where $\phi(\mathrm{x})$ is the standard normal cumulative distribution function *(cdf)*. In communications, $X$ usually represents power variation measured in dB. Considering $\mathrm{X}_{dB}$ with mean $\mu_{dB}$ and variance $\sigma_{dB}^2$, the corresponding lognormal RV $L = e^{\xi \cdot X_{dB}} = 10^{\frac{X_{dB}}{10}}$ has the following pdf:

$$f_L(l, \mu_{dB}, \sigma_{dB}) = \begin{cases} \dfrac{1}{\sqrt{2\pi} l \sigma_{dB}} \exp(-\dfrac{1}{2\sigma_{dB}^2}[10\log(l) - \mu_{dB}]^2) & l > 0 \\ 0 & otherwise \end{cases} \qquad (2)$$

Where
$$\xi = \frac{\ln(10)}{10}$$
and
$$\mu_X = \xi \mu_{dB}$$
$$\sigma_X^2 = \xi^2 \sigma_{dB}^2$$

Lognormal sum distribution corresponds to the sum of independent lognormal RVs, i.e.

$$\Lambda = \sum_{i=1}^{N} L_n = \sum_{i=1}^{N} e^{X_n} \qquad (3)$$

## 2.2. Lognormal sum moments computation

It is well known that lognormal sum distribution function has no closed form, but it is possible to derive its moment. Let $L_i$ be a lognormal RV with mean $m_i$ and variance $D_i^2$. Let $\mu_i, \sigma_i^2$ be the mean and the variance of the corresponding normal distribution.

To compute $m_i, D_i^2$, the $p$th moment $\alpha_{i,p}$ about the origin is first calculated for a lognormal distribution :

$$\alpha_{i,p} = \int_0^{+\infty} t^p \frac{1}{\sqrt{2\pi}} e^{-(\ln(t) - \mu_i)^2 / 2\sigma_i^2} dt \qquad (4)$$

With the substitution:

$$z = \frac{\ln(t) - \mu_i}{\sigma_i}, \qquad \frac{dz}{dt} = \frac{dt}{t\sigma_i}, \qquad t = e^{\sigma_i z + \mu_i}$$

The $p$th moment $\alpha_{i,p}$ of RV $L_i$ is:

$$\alpha_{i,p} = e^{\mu_i p} \int_{-\infty}^{+\infty} e^{\sigma_i p z} \frac{1}{\sqrt{2\pi}} e^{-z^2/2} dz \qquad (5)$$

With $\sigma_i p = \hat{i} w$, the integral reduces to the characteristic function of the standard normal distribution, which has the value $e^{-w^2/2}$, so that:

$$\alpha_{i,p} = e^{\mu_i p} e^{\sigma_i^2 p^2 / 2} \qquad (6)$$

Then, we get

$$m_i = e^{\mu_i} e^{\sigma_i^2 / 2} \qquad (7)$$

$$D_i^2 = \alpha_{i,2} - \alpha_{i,1}^2 = e^{2\mu_i} e^{\sigma_i^2} (e^{\sigma_i^2} - 1) \qquad (8)$$

Based on lognormal moment's computation, we derive the first two central moment of the sum of lognormal distribution:

$$m = \sum_{i=1}^{N} m_i = \sum_{i=1}^{N} e^{\mu_i} e^{\sigma_i^2 / 2} \qquad (9)$$

$$D^2 = \sum_{i=1}^{N} D_i^2 = \sum_{i=1}^{N} e^{2\mu_i} e^{\sigma_i^2} (e^{\sigma_i^2} - 1) \qquad (10)$$

## 2.3. Tails properties of Lognormals sum

SLN distribution has linear asymptotes with simple closed–form expressions in both tails. To study tails behaviour of lognormal sum, it is convenient to work on lognormal probability scale [6], i.e., under the transformation G:

$$G : F(\mathrm{x}) \mapsto \widetilde{F}(\mathrm{x}) = F(\phi^{-1}(F(e^x))) \qquad (11)$$

Under this transformation, the lognormal distribution is mapped onto a linear equation. It has been shown in [20] that the slope of the right tail of the SLN *cdf* on lognormal probability scale is equal to $1/\max_i\{\sigma_i\}$.

$$\lim_{x \to +\infty} \frac{\delta}{\delta x} \widetilde{F}_{SLN}(x) = \frac{1}{\max_i\{\sigma_i\}} \qquad (12)$$

In [21] Szyszkowicz and Yanikomeroglu argued that the slope of the left tail of the SLN *cdf* on lognormal probability scale is equal to $\sqrt{\sum_{i=1}^{N} \sigma_i^{-2}}$. A result which has been proved formally by Gulisashvili and Tankov in [22].

$$\lim_{x \to -\infty} \frac{\delta}{\delta x} \widetilde{F}_{SLN}(x) = \sqrt{\sum_{i=1}^{N} \sigma_i^{-2}} \qquad (13)$$

# 3. LOG SKEW NORMAL DISTRIBUTION

## 3.1. Log Skew Normal PDF

The standard skew normal distribution appeared firstly in [26] and was independently proposed and systematically investigated by Azzalini [27]. The random variable X is said to have a scalar $SN(\lambda, \varepsilon, \omega)$ distribution if its density is given by:

$$f_X(x; \lambda, \varepsilon, \omega) = \frac{2}{\omega} \varphi(\frac{x-\varepsilon}{\omega}) \phi(\lambda \frac{x-\varepsilon}{\omega}) \qquad (14)$$

Where

$$\varphi(\mathrm{x}) = \frac{e^{-x^2/2}}{\sqrt{2\pi}}, \qquad \phi(\mathrm{x}) = \int_{-\infty}^{x} \varphi(\zeta) \, d\zeta$$

With $\lambda$ is the shape parameter which determines the skewness, $\varepsilon$ and $\omega$ represent the usual location and scale parameters and $\varphi$, $\phi$ denote, respectively, the *pdf* and the *cdf* of a standard Gaussian RV.

The cdf of the skew normal distribution can be easily derived as:

$$F_X(x; \lambda, \varepsilon, \omega) = \phi(\frac{x-\varepsilon}{\omega}) - 2\,T(\frac{x-\varepsilon}{\omega}, \lambda) \qquad (15)$$

Where function $T(x, \lambda)$ is Owen's *T* function expressed as:

$$T(x, \lambda) = \frac{1}{2\pi} \int_0^\lambda \frac{\exp\left\{-\frac{1}{2}x^2(1+t^2)\right\}}{(1+t^2)} dt \qquad (16)$$

Similar to the relation between normal and lognormal distributions, given a skew normal RV X then $L = e^{\xi X} = 10^{\frac{X}{10}}$ is a log skew normal distribution. The cdf and pdf of $L$ can be easily derived as:

$$f_L(l; \lambda, \varepsilon, \omega) = \begin{cases} \dfrac{2}{\omega l} \varphi(\dfrac{\ln(l)-\varepsilon}{\omega}) \phi(\lambda \dfrac{\ln(l)-\varepsilon}{\omega}) & l > 0 \\ 0 \quad otherwise \end{cases} \qquad (17)$$

$$F_L(l; \lambda, \varepsilon, \omega) = \begin{cases} \phi(\dfrac{\ln(l)-\varepsilon}{\omega}) - 2T(\dfrac{\ln(l)-\varepsilon}{\omega}, \lambda) & l > 0 \\ 0 \quad otherwise \end{cases} \qquad (18)$$

### 3.2. Log Skew Normal moments computation

In [18], computation of log skew normal moments is done in logarithm domain, then parameters $\lambda, \varepsilon$ and $\omega$ are derived by matching the moment in logarithm domain which require Monte Carlo simulation. In this section we propose to derive the expression of moments of log skew normal distribution (in linear domain), then we can use the moment matching in linear domain to find optimal distribution parameters.

Let $L$ be a log skew normal RV with mean $\xi$ and variance $\varpi^2$, $\varepsilon, \omega^2$ mean and variance of the corresponding skew normal distribution.

To compute $\xi, \varpi^2$, the $p$th moment $\alpha_p$, about the origin is first calculated for a log skew normal distribution :

$$\alpha_p = \int_0^{+\infty} t^p \frac{2}{\omega t} \varphi(\frac{\ln(t)-\varepsilon}{\omega}) \phi(\lambda \frac{\ln(t)-\varepsilon}{\omega}) dt \qquad (19)$$

Using same variable substitution as in (4):

$$z = \frac{\ln(t)-\varepsilon}{\omega} \ , \quad dz = \frac{dt}{t\omega} \ , \qquad t = e^{\omega z + \varepsilon}$$

The $p$th moment $\alpha_p$ of RV $L$ is:

$$\alpha_p = 2e^{\varepsilon p} \int_{-\infty}^{+\infty} e^{\omega pz} \varphi(z) \phi(\lambda z) \ dz$$
$$= e^{\varepsilon p} M_X(p\omega) \qquad (20)$$

In [27], Azzalini showed that the *MGF* of the skew normal distribution is:

$$M_X(t) = E\left[e^{tX}\right]$$

$$= 2e^{t^2/2}\phi(\beta\, t), \qquad \beta = \frac{\lambda}{\sqrt{1+\lambda^2}} \tag{21}$$

Then the pth moment $\alpha_p$ can be expressed as:

$$\alpha_p = e^{\varepsilon p}\, M_X(p\omega)$$

$$= 2e^{\varepsilon p}e^{\omega^2 p^2/2}\phi(\beta\omega p), \qquad \beta = \frac{\lambda}{\sqrt{1+\lambda^2}} \tag{22}$$

Then

$$\xi = 2e^{\varepsilon}e^{\omega^2/2}\phi(\beta\omega) \tag{23}$$

$$\varpi^2 = 2e^{2\varepsilon}e^{\omega^2}(e^{\omega^2}\phi(2\beta\omega) - 2\phi^2(\beta\omega)) \tag{24}$$

## 3.3. Tail properties of Skew Log Normal

In [21] Szyszkowicz and Yanikomeroglu defined the concept of *best lognormal fit* to a tail i.e. that the approximating distribution function have a similar behavior to lognormal sum at a given tail. It is possible to show that LSKN has a best lognormal fit at both tails (see Appendix A). In [23] Capitanio showed that the rate of decay of the right tail probability of a skew normal distribution is equal to that of a normal variate, while the left tail probability decays to zero faster. This result has been confirmed by Hürlimann in [24]. Based on that, it is easy to show that the rate of decay of the right tail probability of a log skew normal distribution is equal to that of a lognormal variate.

Under the transformation G, skew lognormal distribution has a linear asymptote in the upper limit with slope

$$\lim_{x\to+\infty}\frac{\delta}{\delta x}\widetilde{F}_{LSKN}(x) = \frac{1}{w} \tag{25}$$

In the lower limit, it has no linear asymptote, but does have a limiting slope

$$\lim_{x\to-\infty}\frac{\delta}{\delta x}\widetilde{F}_{LSKN}(x) = \frac{\sqrt{1+\lambda^2}}{w} \tag{26}$$

These results are proved in Appendix A. Therefore, it will be possible to match the tail slopes of the LSKN with those of the SLN distribution in order to find LSKN optimal parameters.

## 3.4. Moments and lower tail slope matching

It is possible to extend moment matching method in order to include the third and fourth central moment; however by simulation it seems that higher moments introduce errors to computed parameters. Also upper slope tail match is valid only for the sum of high number of lognormal RVs. In order to derive log skew normal optimal parameters, on consider only the first two central moment and lower slope tail match. We define $\lambda_{opt}$ as solution the following nonlinear equation:

$$\frac{\sum_{i=1}^{N} e^{2\mu_i} e^{\sigma_i^2} (e^{\sigma_i^2}-1)}{(\sum_{i=1}^{N} e^{\mu_i} e^{\sigma_i^2/2})^2} = e^{\frac{1+\lambda^2}{\sum_{i=1}^{N} \sigma_i^{-2}}} \frac{\phi(2\frac{\lambda}{\sqrt{\sum_{i=1}^{N} \sigma_i^{-2}}})}{2\phi^2(\frac{\lambda}{\sqrt{\sum_{i=1}^{N} \sigma_i^{-2}}})} - 1 \qquad (27)$$

Moments and lower tail slope matching leads us to a nonlinear equation which can be solved using different mathematical utility such as "fsolve" in Matlab. Such nonlinear equation need a starting solution guess to converge rapidly, we propose to use upper tail slope match for optimal starting guess solution:

$$\lambda_0 = \sqrt{\max_i (\sigma_i)^2 \sum_{i=1}^{N} \sigma_i^{-2} - 1} \qquad (28)$$

Optimal location and scale parameters $\varepsilon_{opt}, \omega_{opt}$ are obtained according to $\lambda_{opt}$.

$$\begin{cases} \omega_{opt} = \sqrt{\dfrac{1+\lambda_{opt}^2}{\sum_{i=1}^{N} \sigma_i^{-2}}} \\ \varepsilon_{opt} = \ln(\sum_{i=1}^{N} e^{\mu_i} e^{\sigma_i^2/2}) - \dfrac{\omega_{opt}^2}{2} - \ln(\phi(\dfrac{\lambda_{opt}}{\sqrt{\sum_{i=1}^{N} \sigma_i^{-2}}})) \end{cases} \qquad (29)$$

# 4. SIMULATION RESULTS

In this section, we examine the results of the proposed matching method and compare with other priori methods (lognormal approximation, MPLN approximation) in some cases. Monte Carlo simulation results are used as reference.

Table I. compares calculated values of LSKN parameters for different case of sum of lognormal RVs using proposed method and Monte Carlo simulation. It is obvious that our fitting method performs well especially in case of low value of standard deviation.

Fig. 1 and Fig. 2 show the results for the cases of the sum of 20 lognormal RVs with mean 0dB and standard deviation 3dB and 6dB. The CDFs are plotted in lognormal probability scale [6]. Simulation results show that the accuracy of our approximation get better as the number of lognormal distributions increase. We can see that LSKN approximation offers accuracy over the entire body of the distribution for both cases.

Table 1.       LSKN parameters for different cases of sum of Lognormals

| Case of Sum of lognormal RVs | Proposed Method | | | Monte Carlo Simulation | | |
|---|---|---|---|---|---|---|
| | $\beta$ | $\varepsilon$ | $\omega$ | $\beta$ | $\varepsilon$ | $\omega$ |
| 20 RVs, $\mu$ =0dB, $\sigma$ =3dB (Fig. 1) | 0.6332 | 3.1186 | 0.1996 | 0.6113 | 3.1231 | 0.1975 |
| 20 RVs, $\mu$ =0dB, $\sigma$ =6 dB (Fig. 2) | 0.8749 | 3.3937 | 0.6379 | 0.8776 | 3.4186 | 0.6121 |
| 12 RVs, $\mu$ =[-12...12]dB, $\sigma$ =6 dB (Fig. 3) | 0.9344 | 3.5285 | 1.1194 | 0.8921 | 3.6402 | 1.0441 |
| 6 RVs, $\mu$ =0dB, $\sigma$ =[1..6]dB (Fig. 4) | 0.9766 | 1.3882 | 0.8775 | 0.9933 | 1.4843 | 0.7889 |

In Fig. 3, we consider the sum of 12 lognormal RVs having the same standard deviation of 3dB, but with different means. It is clear that LSN approximation catch the entire body of SLN distribution. In this case, both LSKN and MPLN provides a tight approximation to SLN distribution. However LSKN approximation outperform MLPN approximation in lower region. Since interferences modeling belongs to this case (i.e. same standard deviation with different means), it is important to point out that log skew normal distribution outperforms other methods in this case.

Fig. 4 show the case of the sum of 6 lognormal RVs having the same mean 0dB but with different standard deviations. We can see that Fenton-Wilkinson approximation method can only fit a part of the entire sum of distribution, while the MPLN offers accuracy on the left part of the SLN distribution. However, it is obvious that LSKN method provides a tight approximation to the SLN distribution except a small part of the right tail.

It is worthy to note that in all cases, log skew normal distribution provide a very tight approximation to SLN distribution in the region of CDF less than 0.999.

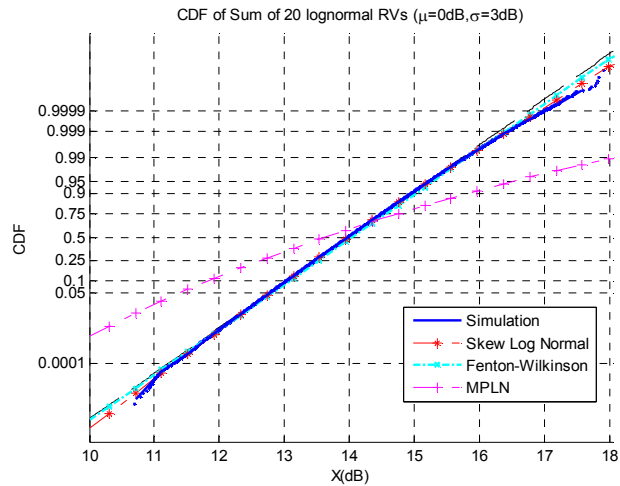

Figure 1. CDF of a sum of 20 i.i.d. lognormal RVs with $\mu$ = 0 dB and $\sigma$ = 3dB.

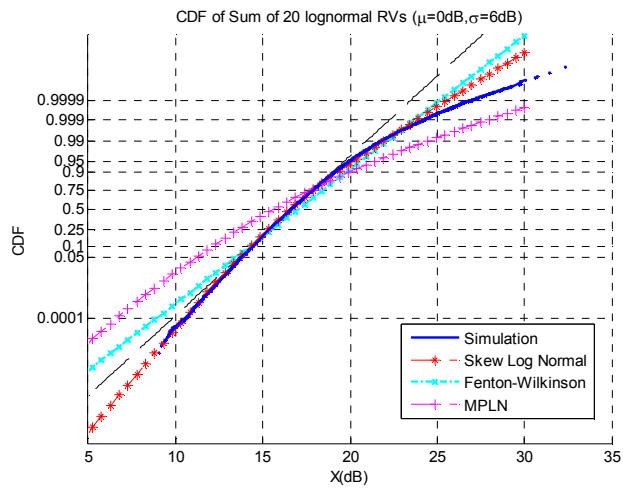

Figure 2. CDF of a sum of 20 i.i.d. lognormal RVs with $\mu = 0$ dB and $\sigma = 6$dB.

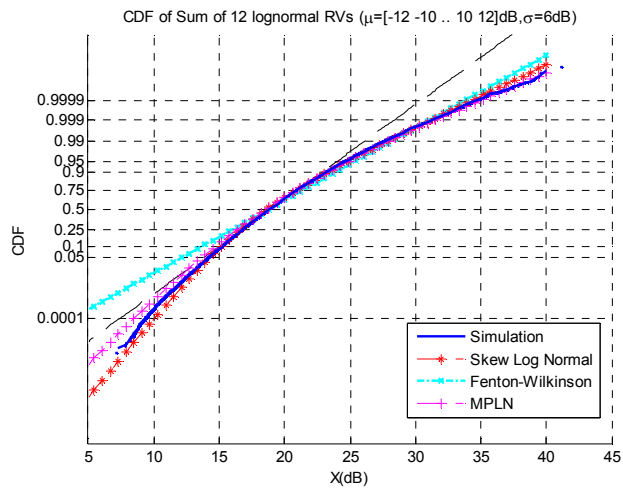

Figure 3. CDF of a sum of 12 lognormal RVs with $\mu = [-12 \ -10 \ -8 \ ... \ 8 \ 10 \ 12]$ dB and

$\sigma = 6$dB.

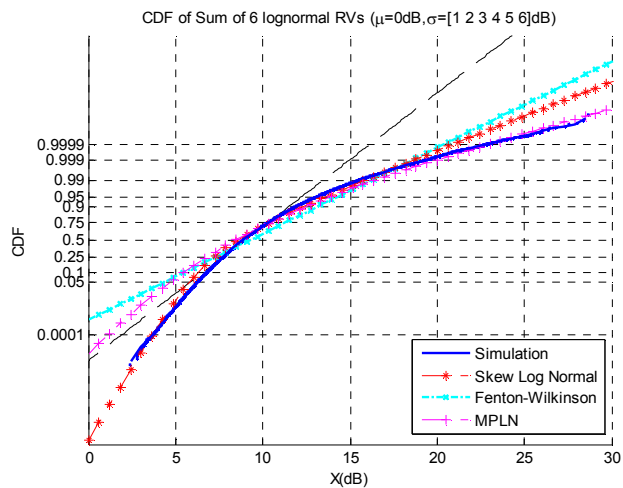

Figure 4. CDF of a sum of 6 lognormal RVs with $\mu = 0$ dB and $\sigma = [1 \ 2 \ 3 \ 4 \ 5 \ 6]$ dB.

# 5. APPLICATION: OUTAGE PROBABILITY IN LOGNORMAL SHADOWING ENVIRONMENT

In this section, we provide an example for outage probability calculation in lognormal shadowing environment based on log skew normal approximation.

We consider a homogeneous hexagonal network made of 18 rings around a central cell. Fig. 5 shows an example of such a network with the main parameters involved in the study: R, the cell range (1 km), Rc, the half-distance between BS. We focus on a mobile station (MS) $u$ and its serving base station (BS), $BS_i$, surrounded by M interfering BS

To evaluate the outage probability, the noise is usually ignored due to simplicity and negligible amount. Only inter-cell interferences are considered. Assuming that all BS have identical transmitting powers, the SINR at the $u$ can be written in the following way:

$$SINR = \frac{S}{I+N} = \frac{P_i K r_{i,u}^{-\eta} . Y_{i,u}}{\sum_{j=0,j\neq i}^{M} P_j K r_{j,u}^{-\eta} . Y_{j,u} + N} = \frac{r_i^{-\eta} . Y_{i,u}}{\sum_{j=0,j\neq i}^{M} r_j^{-\eta} . Y_{j,u}} \qquad (30)$$

The path-loss model is characterized by parameters K and $\eta > 2$. The term $P_i K r_i^{-\eta}$ is the mean value of the received power at distance $r_i$ from the transmitter $BS_i$. Shadowing effect is represented by lognormal random variable $Y_{i,u} = 10^{\frac{x_{i,u}}{10}}$ where $x_{i,u}$ is a normal RV, with zero mean and standard deviation σ, typically ranging from 3 to 12 dB.

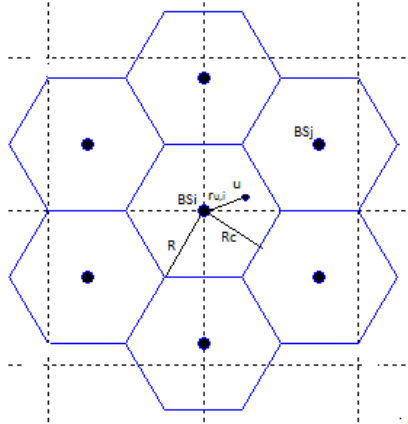

Figure 5. Hexagonal network and main parameters

The outage probability is now defined as the probability for the $\gamma$ SINR to be lower than a threshold value $\delta$ :

$$P(\gamma < \delta) = P(\frac{P_{\text{int}}}{P_{ext}} < \delta) = P(\frac{r_i^{-\eta}.Y_{i,u}}{\sum\limits_{j=0,\,j \neq i}^{M} r_j^{-\eta}.Y_{j,u}} < \delta)$$

$$= P(10\log(\frac{r_i^{-\eta}.Y_{i,u}}{\sum\limits_{j=0,\,j \neq i}^{M} r_j^{-\eta}.Y_{j,u}}) < \delta_{dB}) \qquad (31)$$

$$= P(P_{\text{int},dB} - P_{ext,dB} < \xi \delta_{dB})$$

Where:

$P_{\text{int},dB} = \ln(r_i^{-\eta}.Y_{i,u})$ a normal RV with mean $m = \ln(r_i^{-\eta})$ and standard deviation $\xi\sigma$ .

$P_{\text{ext},dB} = \ln(\sum\limits_{j=0,\,j \neq i}^{M} r_j^{-\eta}.Y_{j,u})$ a skew normal RV with distribution $SN(\lambda, \varepsilon, \omega)$ .

It is easy to show that the difference $P_{\text{int},dB} - P_{ext,dB}$ has a skew normal distribution $SN(\lambda_1, \varepsilon_1, \omega_1)$

Where

$$\varepsilon_1 = \text{m} - \text{e}$$

$$\omega_1 = \sqrt{\xi^2\sigma^2 + \omega^2}$$

$$\lambda_1 = \frac{\lambda}{\sqrt{(1+\lambda^2)(\frac{\xi^2\sigma^2}{\omega^2})+1}}$$

Fig. 6 and Fig. 7 show the outage probability at cell edge (r=Rc) and inside the cell (r=Rc/2), resp. for =3dB and 6dB assuming $\eta$=3. Difference between analysis and simulation results is less than few tenths of dB. This accuracy confirms that the LSKN approximation considered in this work is efficient for interference calculation process.

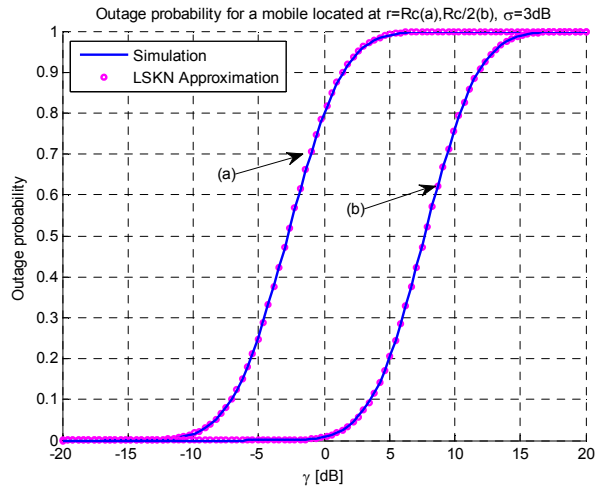

Figure 6. Outage probability for a mobile located at r=Rc (a), r=Rc /2(b), $\sigma$ =3dB, $\eta$=3

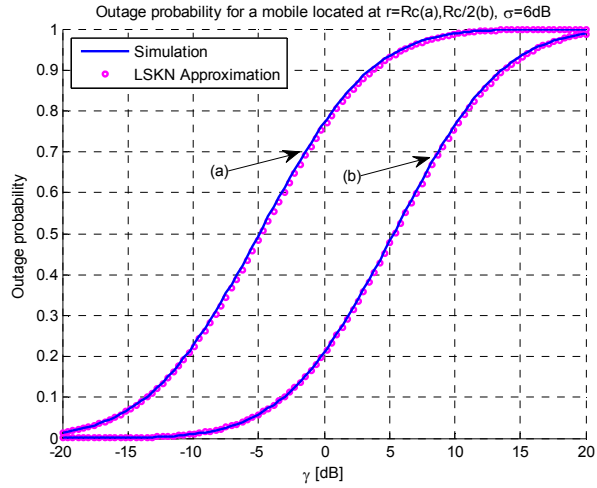

Figure 7. Outage probability for a mobile located at r=Rc (a), r=Rc /2(b), $\sigma$ =6dB, $\eta$=3

## 6. CONCLUSIONS

In this paper, we proposed a fitting method to log skew normal distribution in order to approximate the sum of lognormal distributions. The proposed fitting method uses moment and tails slope matching technique to generate distribution parameters. By obtaining such parameters through a simple procedure, the LSN random variable approximates the entire lognormal sum distribution with high accuracy, especially at the lower region. Simulations confirm that our fitting method outperforms other priori methods for all cases except for the case when lognormal distributions have different standard deviation. Using an example for outage probability calculation in lognormal shadowing environment, we proved that LSKN approximation, considered in this work, is efficient for interference calculation process.

## Appendix A

We begin by showing that the LSKN distribution has a *best lognormal fit* [27] at the right tail. Using [22, Lemma 1], for a given $\lambda > 0$

$$\lim_{x \to +\infty} \frac{1 - F_{LSKN,\lambda}(x)}{1 - \phi(\frac{\ln(x) - \varepsilon}{w})}$$

$$= \lim_{x \to +\infty} \frac{1 - F_{SKN,\lambda}(x)}{1 - \phi(\frac{x - \varepsilon}{w})}$$

$$= \lim_{x \to +\infty} (1 - F_{SKN,\lambda}(x)) \frac{x}{\varphi(x)} = 2$$

Because this ratio converges to a finite non–zero value, we may conclude from [21, Lemma 2] that $\phi(\frac{x - \varepsilon}{w})$ is a *best lognormal fit* to $F_{SKLN}$ at the right tail. The corresponding slope is $1/\omega$ on lognormal paper, which proves (25)

In order to prove the lower tail slope, a few steps are needed. Using [22, Lemma 1], for a given $\lambda > 0$ we have:

$$\lim_{x \to -\infty} F_{SKN,\lambda}(x)$$

$$= \lim_{x \to -\infty} (1 - F_{SKN,-\lambda}(-x))$$

$$= \lim_{x \to +\infty} (1 - F_{SKN,-\lambda}(x))$$

$$= \lim_{x \to +\infty} \sqrt{\frac{2}{\pi}} \frac{\varphi(x\sqrt{1+\lambda^2})}{\lambda(1+\lambda^2)x^2}$$

$$= \lim_{x \to +\infty} \frac{e^{\frac{(1+\lambda^2)x^2}{2}}}{\pi\lambda(1+\lambda^2)x^2}$$

Using approximation from [25]:

$$\phi^{-1} \xrightarrow{x \to 0^+} -\sqrt{-2\ln(2x) - \ln(-\pi\ln(2x))}$$

After some manipulation, we may write:

$$\lim_{x \to -\infty} \frac{\delta}{\delta x} \widetilde{F}_{LSKN}(x) = \frac{1}{\omega} \lim_{x \to -\infty} \frac{\delta}{\delta x} \phi^{-1}(F_{LSKN}(x))$$

$$= -\frac{1}{\omega} \lim_{x \to -\infty} \frac{\delta}{\delta x} \sqrt{(1+\lambda^2)x^2 + o(\ln(-x))}$$

$$= -\frac{1}{\omega} \lim_{x \to -\infty} \frac{2(1+\lambda^2)x}{2\sqrt{(1+\lambda^2)x^2}}$$

$$= \frac{\sqrt{(1+\lambda^2)}}{\omega}$$

Which proves (26).

**Authors**

**Marwane Ben Hcine** was born in Kébili, Tunisia, on January 02, 1985. He graduated in Telecommunications Engineering, from The Tunisian Polytechnic School (TPS), July 2008. In June 2010, he received the master's degree of research in communication systems of the Higher School of Communication of Tunisia (Sup'Com). Currently he is a Ph.D. student at the Higher School of Communication of Tunisia. His research interests are network design and dimensioning for LTE and beyond Technologies.

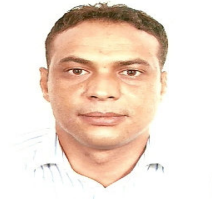

**Pr. Ridha BOUALLEGUE** was born in Tunis, Tunisia. He received the M.S degree in Telecommunications in 1990, the Ph.D. degree in telecommunications in 1994, and the HDR degree in Telecommunications in 2003, all from National School of engineering of Tunis (ENIT), Tunisia. Director and founder of National Engineering School of Sousse in 2005. Director of the School of Technology and Computer Science in 2010. Currently, Prof. Ridha Bouallegue is the director of Innovation of COMmunicant and COoperative Mobiles Laboratory, INNOV'COM Sup'COM, Higher School of Communication. His current research interests include mobile and cooperative communications, Access technique, intelligent signal processing, CDMA, MIMO, OFDM and UWB systems.

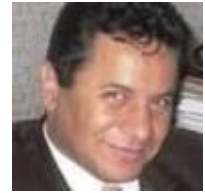